\input{aipcheck}

\documentclass[
    ,final            
  ]
  {aipproc}

\layoutstyle{8x11single}
\newcommand{\aap}{Astron. Astrophys.}
\newcommand{\apj}{Astrophys. J.}
\newcommand{\apjl}{Astrophys. J. Lett.}

\newcommand{\mnras}{Mon. Not. R. Astron. Soc.}
\newcommand{\planss}{Planetary Space Science}

\newcommand{\pra}{Phys. Rev. A}

\newcommand{\prl}{Phys. Rev. Lett.}

\begin{document}

\title{Magnetic Reconnection in Turbulent Plasmas and Gamma Ray Bursts}

\classification{<Replace this text with PACS numbers; choose from this list:
                \texttt{http://www.aip..org/pacs/index.html}>}

\author{A. Lazarian}{
  address={University of Wisconsin-Madison, USA}
}

\author{Huirong Yan}{
  address={KIAA, Peking University, China}
}

\begin{abstract}
 We discuss how the model of magnetic reconnection in the presence of turbulence proposed in
 Lazarian \& Vishniac 1999 makes the reconnection rate independent either of resistivity or microscopic plasma
 effects, but determined entirely by the magnetic field line wandering induced by turbulence. We explain that
 the model accounts for both fast and slow regimes of reconnection and that this property naturally induces flares of reconnection in low
 beta plasma environments. In addition, we show that the model involves volume reconnection which can convert a substantial
 part of the energy into energetic particles. It is important that the reconnection induces an efficient acceleration of the first order Fermi type. 
 Finally, we relate the properties 
 of the reconnection with the observed properties of gamma ray bursts and provide evidence supporting the explanation of 
 gamma ray bursts based on energy release via reconnection.  
\end{abstract}

\maketitle


\section{1. Introduction}

Magnetic reconnection is an exciting problem of extensive history (see a book by Biskamp 2004 and references therein). 
Free energy associated with reconnection can power explosive events like Solar flares and gamma ray bursts. 
For decades, poor understanding of magnetic reconnection prevented all quantitative attempts to characterize these events.
We believe that the situation is different now due to a substantial improvement of our understanding of magnetic reconnection
in realistic turbulent astrophysical environments\footnote{Turbulence is ubiquitous in astrophysics. A large number of different instabilities drive turbulence in astrophysical plasmas. In many respects
the proper question is not what drives the turbulence in a particular situation, but why some reconnection models ignore the existing turbulence. In most
cases, astrophysical turbulence, e.g. turbulence observed in the interstellar media is not arising from reconnection. However, in some situations, in particular,
in the flares of reconnection, the reconnection itself can drive turbulence.}. In particular, a model of reconnection by Lazarian \& Vishniac (1999, henceforth LV99) obtained
the reconnection rate that depended only on the level of turbulence. As the model was supported by numerical testing in 
Kowal et al. (2009, 2012a) and the theoretical studies of the Lagrangian properties of MHD turbulence in Eyink, Lazarian \& Vishniac (2011),
it became important to work on quantitative evaluations of the predictions of the LV99 theory. These predictions include the flares of
magnetic reconnection that can, for instance, explain solar flares (see LV99, Lazarian \& Vishniac 2009). In addition, LV99 model 
entailed the first order Fermi particle acceleration during the reconnection (de Gouveia dal Pino \& Lazarian 2005). This possibility was 
confirmed numerically e.g. in Kowal et al. (2012b). 

A possibility of LV99 model to explain gamma ray bursts was first explored in Lazarian et al. (2003). This idea was elaborated and the results were compared with in observations in Zhang \& Yan (2011). 

In what follows we briefly summarize  the major features of the LV99 reconnection model in \S 2, consider the flares that this model entails in \S 3, discuss the first order Fermi acceleration that induced by LV99 reconnection in \S 4. We consider gamma ray bursts induced by turbulent reconnection in \S 5.  Our discussion and summary is provided in \S 6. 

\section{2. LV99 model of reconnection}

\subsection{Extending Sweet-Parker model for turbulent fields}

\begin{figure}[!t]
\includegraphics[width=1.0 \columnwidth]{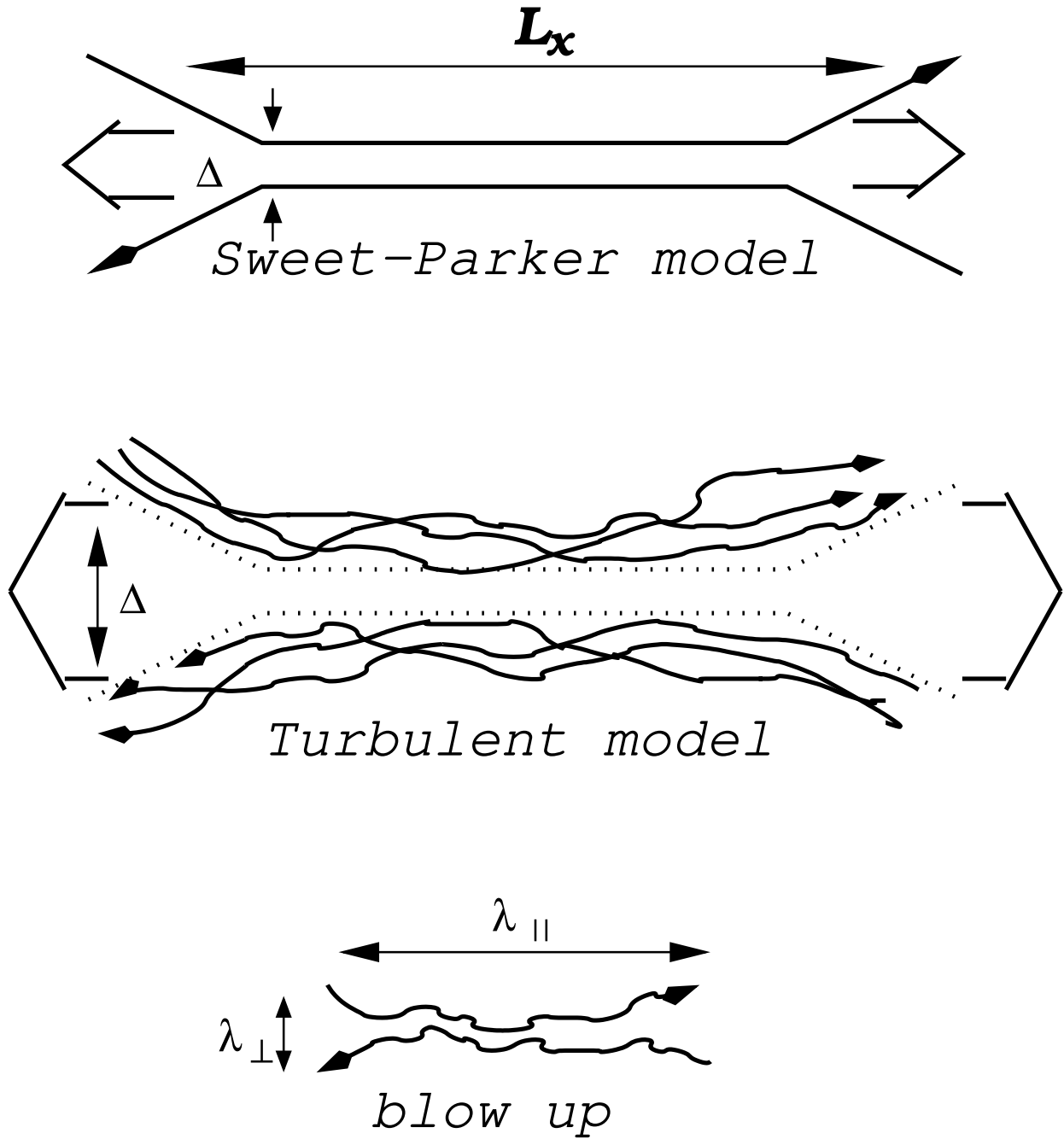}
\caption{
{\it Upper plot}: Sweet-Parker model of reconnection. The outflow is limited by
a thin slot $\Delta$, which is determined by Ohmic diffusivity. The other scale
is an astrophysical scale $L\gg \Delta$.
{\it Middle plot}: Reconnection of weakly stochastic magnetic field according to
LV99. The model that accounts for the stochasticity of magnetic field lines. The
outflow is limited by the diffusion of magnetic field lines, which depends on
field line stochasticity.
{\it Low plot}: An individual small scale reconnection region. The reconnection
over small patches of magnetic field determines the local reconnection rate. The
global reconnection rate is substantially larger as many independent patches
come together. The illustration of LV99 reconnection from Lazarian et al. 2004.
} \label{figure1}
\end{figure}

Reconnection of magnetic field is driven by the free energy of magnetic field and in a generic situation of 3D geometry includes the annihilation of the component of the magnetic field that is different in the interacting magnetic fluxes. The problem of the traditional reconnection approach to 
the reconnection is that the most natural configuration of the reconnection presented by the Sweet-Parker model, which shown in upper part of Figure\ref{figure1} is very slow in astrophysical conditions.
This inefficiency arises from the disparity of the astrophysical scale $L_x$ over which the plasma is being carried into the reconnection region and the
microphysical scale $\Delta$ determined by the plasma resistivity over which the plasma is being ejected from the reconnection region. Taking into
account that the ejection velocity is approximately the Alfven velocity $V_A$ the reconnection rate, 
\begin{equation}
V_{rec}\approx V_A \frac{\Delta}{L_x},
\label{for1}
\end{equation}
is very small, $\ll V_A$. In fact, for the outflow region determined by the Ohmic resistivity $\Delta \approx \eta/V_{rec}$ one recovers the
Sweet-Parker formulae for the reconnection rate $V_{rec, SP}\approx V_A Rm^{-1/2}$, where $Rm\equiv L_xV_A/\eta$, the Lundquist number, can be huge, e.g. of the order $10^{10}$ or even $10^{20}$ for many astrophysical situations. As a result, the reconnection rate in the 
classical Sweet-Parker model is negligible. 

The situation changes dramatically in the presence of turbulence as it shown in the middle part of Figure\ref{figure1}. Studying reconnection in 
the presence of turbulence LV99 showed that the outflow region is determined by the field wandering of magnetic field. Wandering or meandering
of magnetic field is well known (see Jokipii 1973) and numerically tested (see Lazarian, Vishniac \& Cho 2004) effect. This effect has been used for
decades to explain the perpendicular diffusion of cosmic rays in astrophysical magnetic fields. 

To obtain the rate of magnetic field wandering LV99 had to extend the Goldreich-Sridhar (1995, henceforth GS95) model of strong MHD turbulence for the case of
subAlfvenic turbulence\footnote{In addition, LV99 was the first paper which discussed the local system of turbulence which is one of the
key ingredients of the present day understanding of MHD turbulence. The critical balance condition that is the corner stone of strong MHD turbulence is valid
only in the local system of reference. }  In doing so, LV99 obtained the expressions of the magnetic reconnection for subAlfvenic turbulence:
\begin{equation}
V_{rec}\approx V_A\min\left[\left({L_x\over L}\right)^{1/2},
\left({L\over L_x}\right)^{1/2}\right] M_A^2,
\label{recon}
\end{equation}
where $M_A\equiv V_L/V_A$ is the Alfven Mach number and $L$ and $V_L$ are the 
turbulence injection scale and velocity at this scale, respectively.

Naturally, reconnection in the presence of sufAlfvenic turbulence provides the natural generalization of the Laminar Sweet-Parker reconnection. However,
the LV99 reconnection is also applicable to transAlfvenic and superAlfvenic turbulence. The bending of magnetic field lines decreases with the decrease
of the scale and at a sufficiently small scale $l_A=L M_A^{-3}$, the turbulent perturbations get subAlfvenic 
with LV99 model is relevant for the reconnection at this scale and smaller scales.

Finally, we should stress that the LV99 model is independent of small scale physics determining the local reconnection. LV99 conservatively assumed that
the small scale reconnection happens within small scale Sweet-Parker regions (see Figure \ref{figure1}) and showed that this does not present a bottle-neck
of the process. The key of accelerated reconnection is that many independent patches of magnetic flux reconnect simultaneously enhancing the total reconnection
rate. In fact, it is possible to show (see LV99) that even with Sweet-Parker reconnection the overall reconnection rate constrained by the Ohmic diffusion
exceeds $V_A$, which shows that the Ohmic diffusivity does not present any limitation for magnetic reconnection rate. 

LV99 model also considers the motion of the bundles of reconnected flux within the reconnection zone and concludes that their interactions does not present
an impediment for the reconnection rate. This also follows from a more recent paper by Eyink, Lazarian \& Vishniac (2011, henceforth ELV11) where it is shown that magnetic flux is not frozen in turbulent fluids irrespectively of their conductivity. In view of ELV11 the expression given by Eq. (\ref{recon}) is a direct consequence
of this violation of frozen-in condition. This is a result of Richardson diffusion present in turbulent fluids (see Eyink 2011). The possibility of deriving the Eq. (\ref{recon}) from the approach different from the one used in the original LV99 treatment strengthens the theoretical foundations of the discussed extention of Sweet-Parker reconnection in turbulent fluids.

\subsection{Numerical testing of LV99 model}

\begin{figure*}
\includegraphics[trim = 20mm -10mm 20mm -5mm, clip,width=0.4\textwidth]{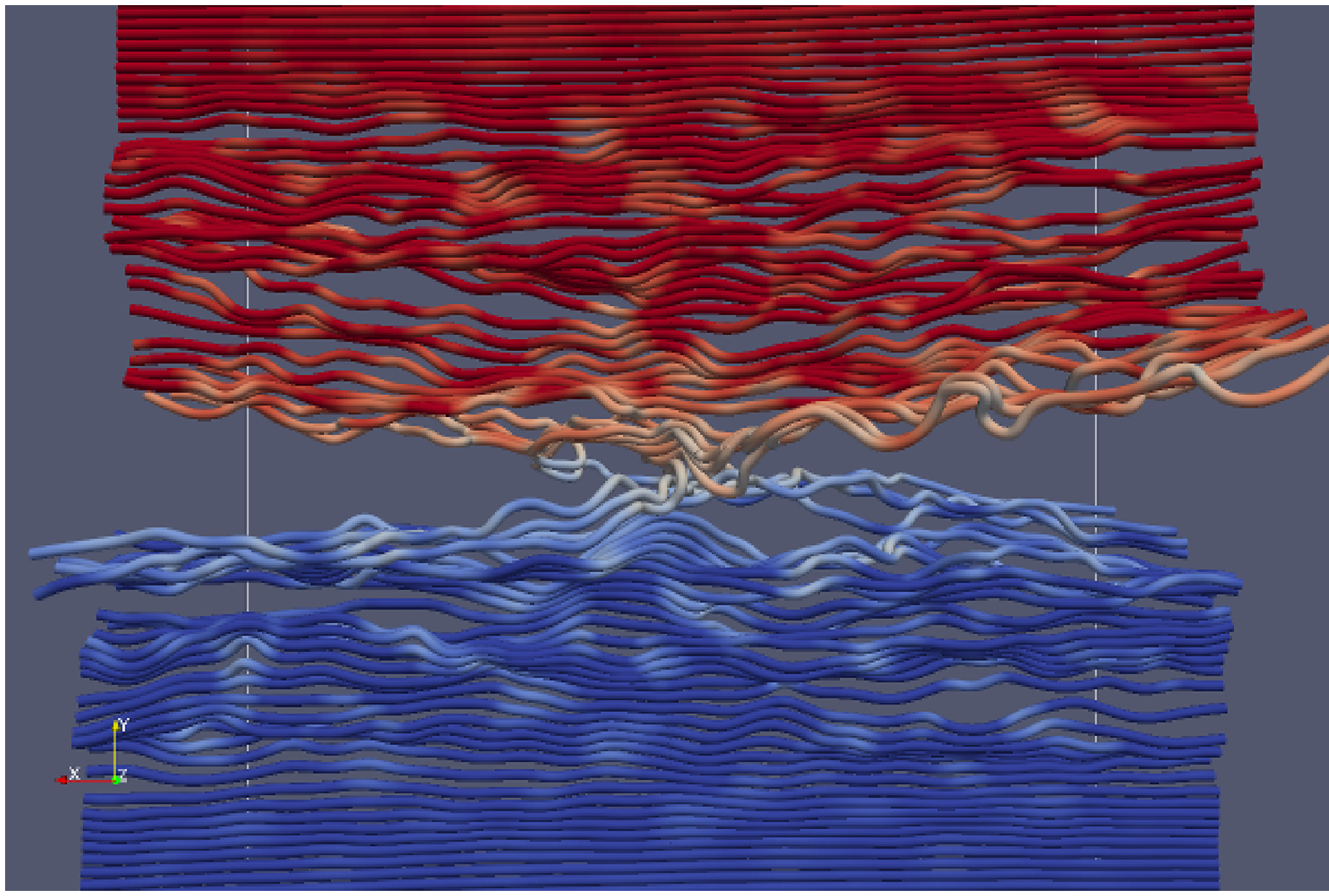}
\includegraphics[width=0.25\textwidth]{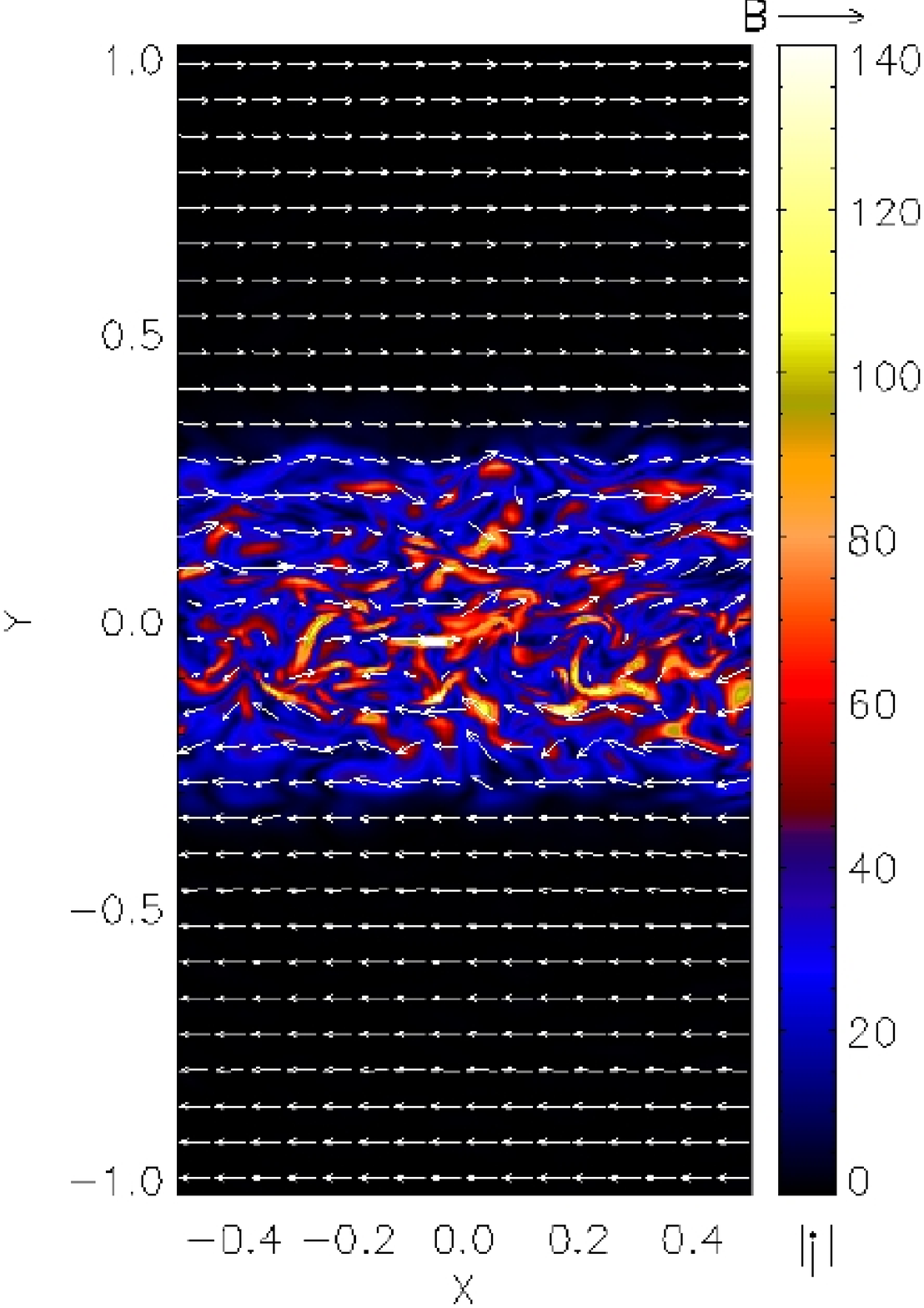}
\includegraphics[width=0.25\textwidth]{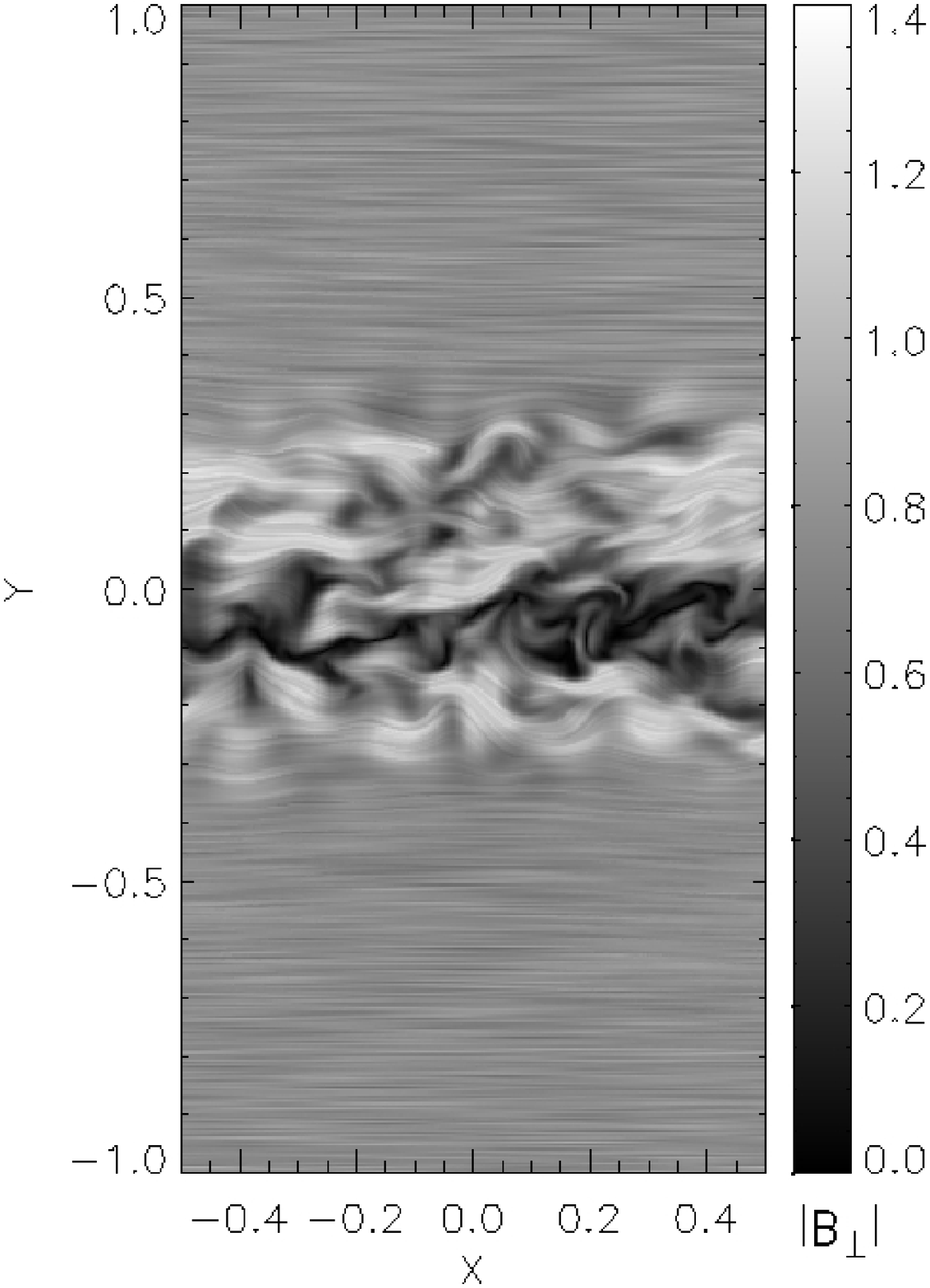}
\caption{ {\it Left panel}: Magnetic field in the reconnection region.  Large
perturbations of magnetic field lines arise from reconnection rather than
driving; the latter is subAlfv\'enic.    The color corresponds to the direction
of magnetic lines.
{\it Central panel}: Current intensity and magnetic field configuration during
stochastic reconnection.  We show a slice through the middle of the
computational box in the xy plane after twelve dynamical times for a typical
run.  The shared component of the field is perpendicular to the page. The
intensity and direction of the magnetic field is represented by the length and
direction of the arrows.  The color bar gives the intensity of the current. The
reversal in $B_x$  is confined to the vicinity of y=0 but the current sheet is
strongly disordered with features that extend far from the zone of reversal.
{\it Right panel}: Representation of the magnetic field in the reconnection zone
with textures. From Lazarian et al. (2011).
\label{figure2}}
\end{figure*}

Numerical studies of reconnection is tricky as the Lundquist numbers corresponding to numerical simulations are puny compared to astrophysical one.
The usual routine of increasing the resolution several times to see the effect on reconnection may not work due to this disparity. For instance, even a weak
dependence of reconnection rate on the Lundquist number, which is extremely difficult to notice with the traditional "change by factor of two" approach, can make
the reconnection rate completely negligible for for astrophysics, as the corresponding numbers increase by a factor of $10^{10}$ or larger. Thus it is essential that LV99
provides analytical dependences (see Eq. (\ref{recon})) that can be tested numerically. 

Numerical studies of LV99 model were presented in Kowal et al. (2009, 2012a). The numerical set-up was used to test 3D magnetic reconnection in the presence of guide field (i.e. magnetic field fluxes intersect at arbitrary angles) and in the presence of subAlfvenic turbulence. The reconnection was {\it not} forced and open boundary conditions were employed. Turbulence was driven both in Fourier space and in real space. Both the input power of the turbulence driving and the scale
of the driving were varied to test the analytical predictions of the LV99 model. 

It is illustrated in Figure \ref{figure2} that the turbulence in the simulations is driven around the reconnection region. This mimics the effects of astrophysical
turbulence that is known to exist in high Reynolds number astrophysical fluids\footnote{Volume driving that we use in the simulations corresponds well the
actual driving in turbulence in astrophysical conditions, where smaller eddies get energy from the larger ones. This driving results in homogeneous
balanced turbulence as opposed to inhomogeneous imbalanced one resulting from the driving induced at the numerical box boundaries. A theoretical
description of such turbulence is currently absent and predictions of the reconnection rates for such turbulence is currently impossible.}. The driving is
subAlfvenic with only gentle perturbations of the magnetic field lines induced. This driving, nevertheless, results in magnetic field wandering which, in agreement
with LV99 model, opens up the outflow region and induces reconnection independent of resistivity.

\begin{figure*}
\includegraphics[width=0.8\textwidth]{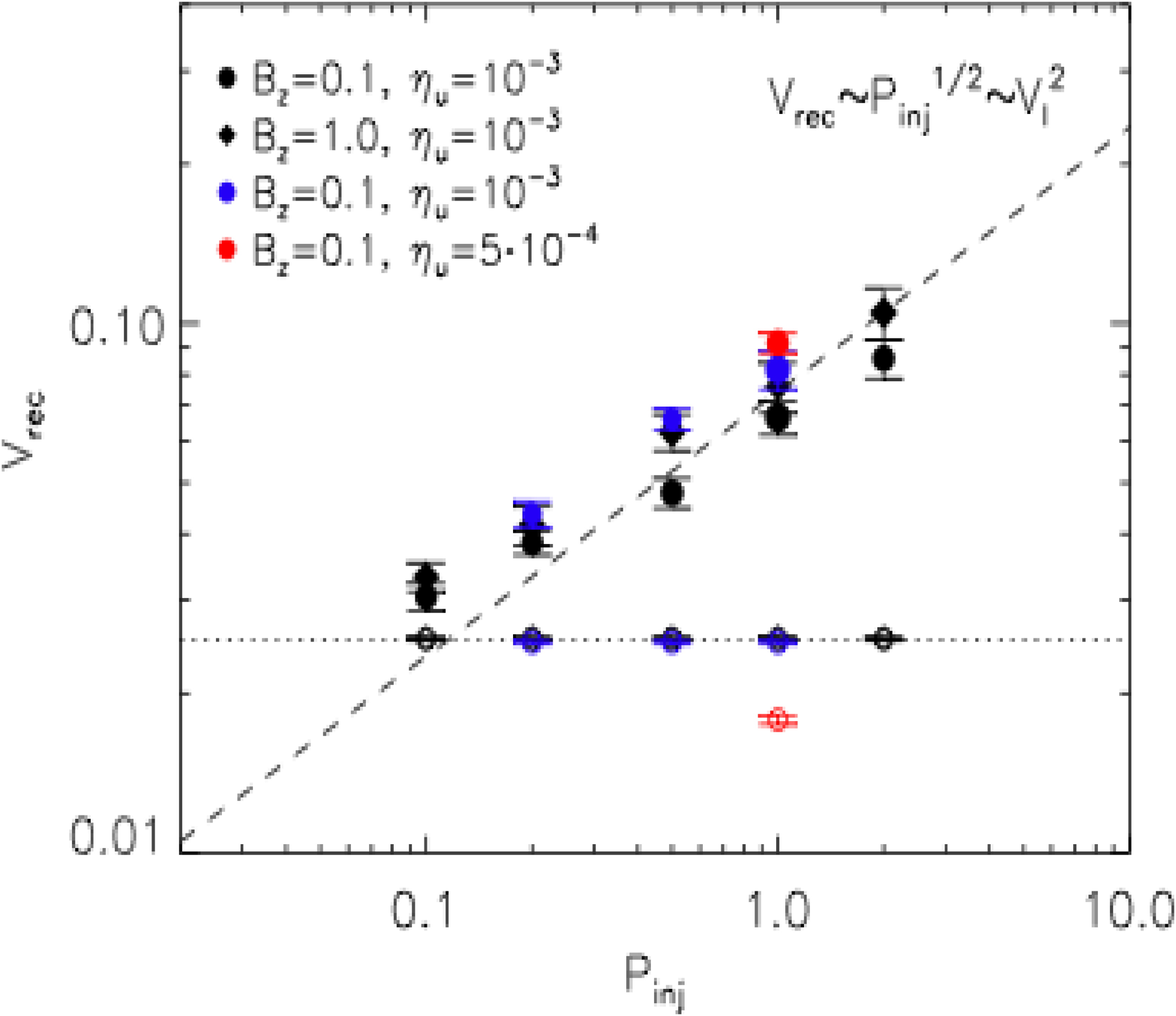}
\caption{Reconnection speed versus input power $P_{inj}$ for the driven turbulence.  {\it
Left}: Variations of reconnection speed in time for different levels of
turbulent driving {\it Right}: Reconnection speed, plotted against the input power for an
injection wavenumber equal to 8 (i.e. a wavelength equal to one eighth of the
box size) and a resistivity $\nu_u$.  The dashed line is a fit to the predicted
dependence of  $P^{1/2}$.  The horizontal line shows the laminar reconnection
rates for each of the simulations before the turbulent forcing started.  From Kowal et al. (2012a).
\label{figure3}}
\end{figure*}

A comparison of results of numerical testing and theoretical predictions obtained in LV99 is shown in Figure \ref{figure3}. 
For subAlfvenic turbulence the input power $P_{inj}$ scales
as $V_L^4$. As a result the predicted reconnection rate (see Eq. \ref{recon}) scales as $V_{rec} \sim P_{inj}^{1/2}$. It is easy to 
see that this dependence is in good agreement with the results of numerical simulations.  In Figure \ref{figure3}
the  uncertainties in the time averages are indicated by the size of the symbols
and the variances are shown by the error bars. The higher resolution 1028x512x512 simulations are shown with red color. Those
corresponds to lower Sweet-Parker rate, but the rate in the presence of turbulence does
not change.

\begin{figure*}
\includegraphics[width=0.4\textwidth]{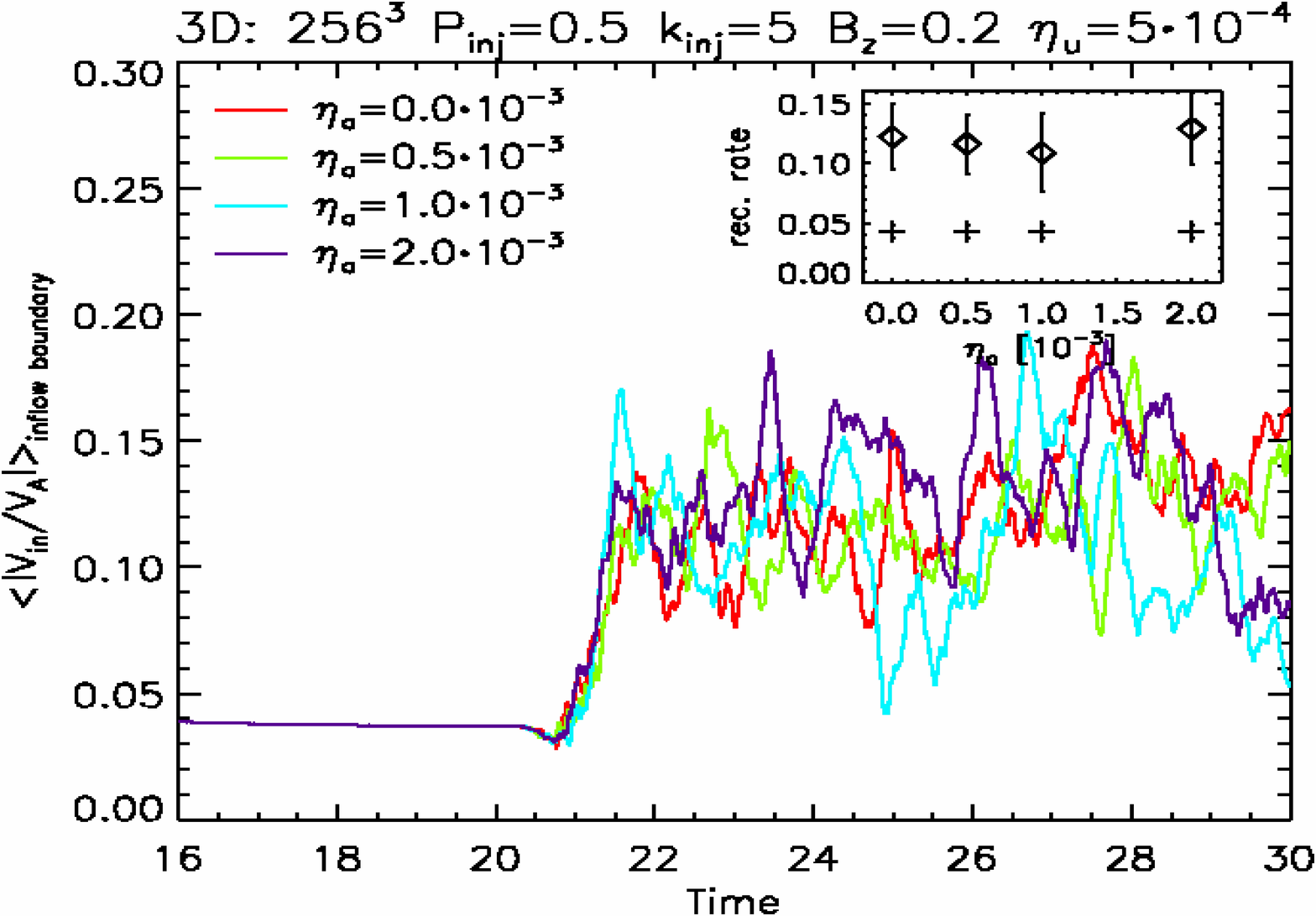}
\includegraphics[width=0.4\textwidth]{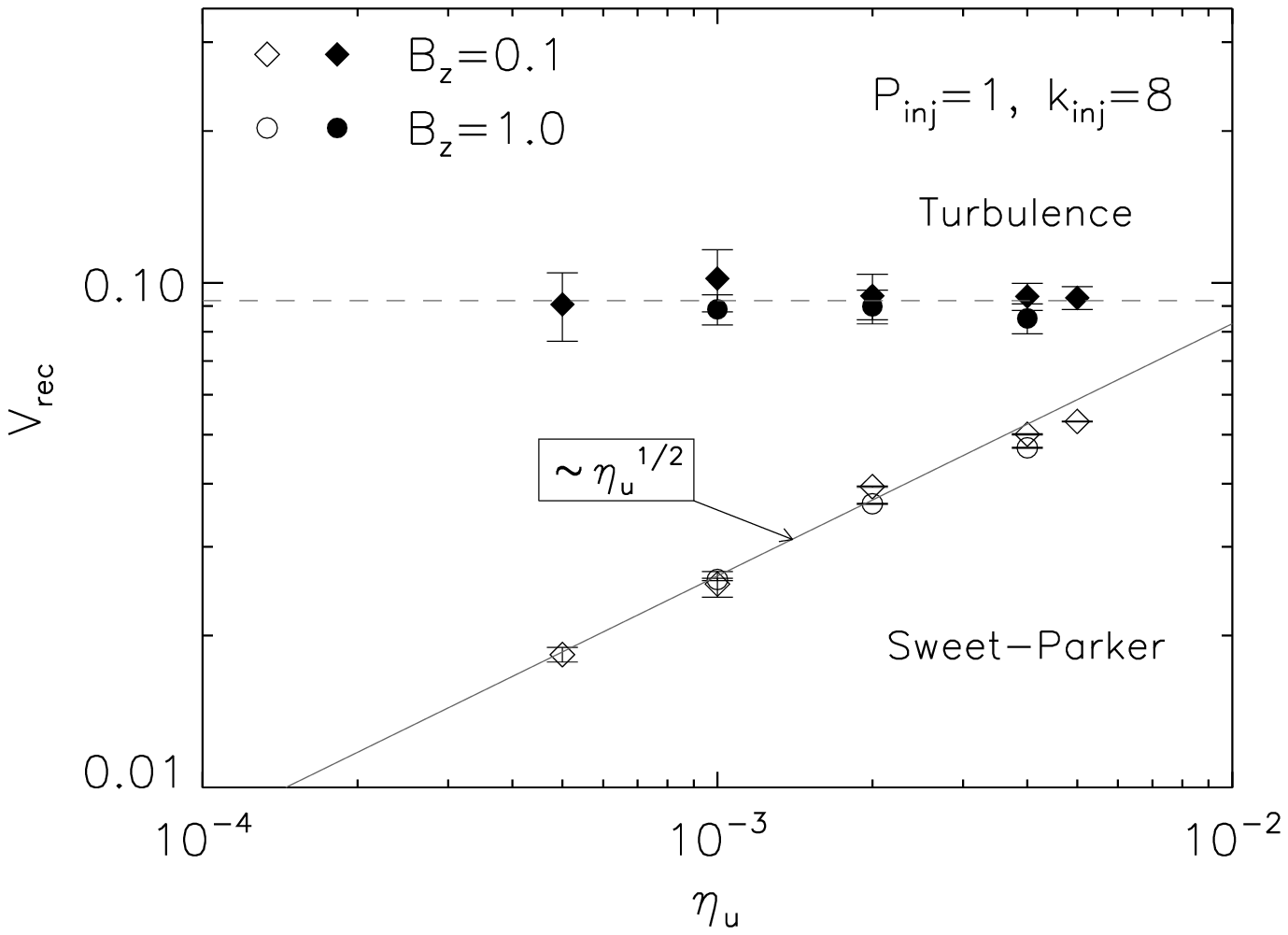}
\caption{{\it Left}: Effect of anomalous resistivity on the reconnection speed. No
dependence is observed. {\it Right}: No dependence on the Ohmic diffusion is observed either. The lower curve reproduces
the Sweet-Parker reconnection rates for the laminar case. From Kowal et al. (2009).
\label{figure4}}
\end{figure*}

For decades the it has been considered that plasma effects, e.g. Hall effect\footnote{A detailed theoretical discussion why 
the Hall term is subdominant in the presence of pre-existing turbulence can be found in ELV11.}, are essential for making magnetic reconnection fast.
The use of anomalous resistivity, i.e. the non-linear resistivity that depends on the value of the current density is one of the
accepted ways to simulate plasma effects. Figure \ref{figure4} shows that the reconnection rate does not change with the
anomalous resistivity in the presence of turbulence. As expected, the reconnection rate of turbulent magnetic field does not depend on the 
Ohmic resistivity either. This confirms that the LV99 reconnection is fast. 

\subsection{Relation to other ideas}

LV99 is a volume reconnection and it was proposed at the time when X-point Petschek reconnection dominated the 
landscape of reconnection ideas (see Shay et al. 1998). Therefore the observations of thick outflow regions was the evidence in favor
of LV99 model. As the publication of the corresponding observational result  (Cirravella \& Raymond 2008) was somewhat delayed,
the competing models evolve towards more resemblance to LV99, getting chaotic structure of magnetic field
and thick outflow regions. This convergence of ideas on reconnection makes it more challenging to distinguish them observationally,
especially, as new modeling indicates not only tearing instability development but a transition to turbulence (see Karimabadi 2012).

The idea of tearing instability (see Loreiro et al. 2007) may be considered to be complementary to the LV99 model. The process,
for instance, can lower the threshold for the reconnection bursts that we discuss in \S 3.

Finally, we mention, that  the LV99 model is radically different from what
was proposed earlier in terms of models of how turbulence can affect reconnection.  
For instance, in a study by Matthaeus \& Lamkin (1985, 1986), which is the closest to the spirit to
 LV99 the authors performed 2D
numerical simulations of turbulence and provided arguments in favor of magnetic
reconnection getting fast. However, the physics of the processes that they
considered was very different from that in LV99.  For instance, the key process
of field wandering of the LV99 model has not been considered in the aforementioned
 papers.  On the contrary,  processes that they discuss, e.g. X-point  reconnection
 and possible effects of heating and compressibility are
not ingredients of the LV99 model.  Other papers discussing turbulence, e.g. Speiser (1970) and
Jacobson (1984) are even more distant in terms of the effects that they
explored. The aforementioned authors studied the changes of the microscopic properties of the
plasma induced by turbulence and considered how these changes can accelerate
magnetic reconnection. On the contrary, LV99 shows that the microscopic plasma
properties are irrelevant for their model of reconnection (see also Figure \ref{figure4}).

\section{3. Flares of Reconnection}
\label{flares}
\subsection{Reconnection instability}

In the text above we have discussed how the pre-existing turbulence changes the nature of reconnection. 
The next question is what happens when the magnetic fields are initially laminar. The answer to this question 
in LV99 is that a {\it reconnection instability} exists and can drive reconnection in a bursty fashion.
Indeed,  These bursts of reconnection are easy to understand by considering low $\beta$ (i.e. highly magnetized)
plasmas with magnetic laminar flux tubes coming into contact with each other. Initially, the
magnetic reconnection is going to proceed at a slow pace (see Figure \ref{figure1}), as magnetic field lines are laminar.
However, the system of two highly magnetized flux tubes in contact is unstable to the development of turbulence.
Indeed, if the outflow gets turbulent, turbulence should spread over field lines of surrounding flux, inducing their
wandering and extending the width of the outflow region $\Delta$ and increasing the energy injection in
the system via the increase of $V_{rec}$. Both factors increase the level of turbulence in the system\footnote{For
instance, the increase of $\Delta$ increases the Reynolds number of the outflow, making the outflow more turbulent.}
inducing a positive feedback leading to an explosion of reconnection.

A characteristic feature of this reconnection instability is that it is finite amplitude instability and therefore it can allow
the accumulation of the flux prior to reconnection. In other words, LV99 model predicts that the reconnection can 
be both fast and slow, which is the necessary requirement of bursty reconnection frequently observed in nature, e.g.
in Solar flares. This process may be related to the bursts of reconnection observed in simulations in Lapenta (2008). 
In addition, LV99 predicted the process of {\it triggered reconnection} when reconnection in one part of the volume sends 
perturbations that initiate reconnection in adjacent volumes. Such process was reported recently in the observations of 
Sych et al. (2009).  

All in all, in the absence of
external turbulence, the original outflow, e.g. originated through tearing
instability (see Loureiro et al. 2007, Bhattacharjee et al. 2009), gets turbulent
and triggers the mechanism described above. This shows that the tearing and
turbulent mechanisms may be complementary\footnote{When turbulence develops the
LV99 mechanism can provide much faster reconnection compared to tearing and
tearing may become a subdominant process. In fact, emerging turbulence may
suppress the tearing instability.}. We believe that such a mechanism can be important
for explaining a wide variety of astrophysical processes ranging from solar flares to gamma ray bursts
as well as bursty reconnection in the pulsar winds.

\subsection{A Simple model}

A simple quantitative model of flares was presented in Lazarian \& Vishniac (2009). It can 
be understood by considering a reconnection region of length $L_x$ and thickness $\Delta$.  The
thickness is determined by the diffusion of field lines, which is in turn
determined by the strength of the turbulence in the volume.  Reconnection will
allow the magnetic field to relax, creating a bulk flow.  However, since
stochastic reconnection is expected to proceed unevenly, with large variations
in the current sheet, we can expect that some unknown fraction of this energy
will be deposited inhomogeneously, generating waves and adding energy to the
local turbulent cascade.  This is a natural outcome of the turbulent flow

We take the plasma density to be approximately uniform
so that the Alfven speed and the magnetic field strength are interchangeable.
The nonlinear dissipation rate for waves is
\begin{equation}
\tau_{nonlinear}^{-1}\sim\max\left[ {k_\perp^2 v_{wave}^2\over k_\|V_A},k_\perp^2 VL\right],
\end{equation}
where the first rate is the self-interaction rate for the waves and the second
is the dissipation rate by the ambient turbulence (see Beresnyak \& Lazarian 2008). 
The important point here is that $k_\perp$ for the waves falls somewhere in the
inertial range of the strong turbulence.  Eddies at that wavenumber will disrupt
the waves in one eddy turnover time, which is necessarily less than $L/V_A$.
The bulk of the wave energy will go into the turbulent cascade before escaping
from the reconnection zone.  

We can therefore simplify our model for the energy budget in the reconnection
zone by assuming that some fraction $\epsilon$ of the energy liberated by
stochastic reconnection is fed into the local turbulent cascade.  The evolution
of the  turbulent energy density per area is
\begin{equation}
{d\over dt}\left(\Delta V^2\right)=\epsilon V_A^2 V_{rec}-V^2\Delta {V_A\over L_x},
\end{equation}
where the loss term covers both the local dissipation of turbulent energy, and
its advection out of the reconnection zone.  Since $V_{rec}\sim v_{turb}$  and
$\Delta\sim L_x(V/V_A)$,  we can rewrite this by defining $M_A\equiv
V/V_A$ and $\tau\equiv L_x/V_A$ so that
\begin{equation}
{d\over d\tau}M_A^3\approx \epsilon M_A-M^3_A.
\end{equation}
If $\epsilon$ is a constant then
\begin{equation}
V\approx V_A\epsilon^{1/2}\left[1-\left(1-{ M_A^2\over\epsilon}\right)e^{-2\tau/3}\right]^{1/2}.
\end{equation}
This implies that the time during which reconnection rate rises to $\epsilon^{1/2}V_A$ is 
comparable to the ejection time from the reconnection region ($\sim L_x/V_A$).
Given that reconnection events in the solar corona seem to be episodic, with
longer periods of quiescence, this indicates that either $\epsilon$ is very small,
for example, dependent on the ratio of the  thickness of the current sheet to
$\Delta$. If it scales as $M_A$
to some power greater than two then initial conditions dominate the early time
evolution.

Another route by which stochastic reconnection might be self-sustaining
would be in the context of a series of topological knots in the magnetic field,
each of which is undergoing reconnection. The problem is sensitive to
geometry, however.  We assume that as each knot undergoes reconnection it releases a
characteristic energy into a volume which has the same linear dimension as the
distance to the next knot.  The density of the energy input into this volume is
roughly $\epsilon V_A^2 V/L_x$, where $\epsilon$ is the efficiency with
which the magnetic energy is transformed into turbulent energy.  We have
\begin{equation}
\epsilon {V_A^2V\over L_x}\sim {v'^3\over L_k},
\end{equation}
where $L_k$ is the distance between knots and $v'$ is the turbulent velocity
created by the reconnection of the first knot.  This process will proceed
explosively if $v'>V$ or
\begin{equation}
V_A^2 L_k\epsilon> V^2 L_x.
\end{equation}
This condition is easy to fulfill.  The bulk motions created by
reconnection will  generate significant turbulence as they interact
with their surrounding, so $\epsilon$ should be of order unity.  Moreover the
length of any current sheet should be at most comparable to the distance to the
nearest distinct magnetic knot.  The implication is that each magnetic
reconnection event will set off its neighbors, boosting their reconnection rates
from $V_L$, set by the environment, to $\epsilon^{1/2}V_A(L_k/L_x)^{1/2}$ (as
long as this is less than $V_A$).  The process will take a time comparable to
$L_x/V_L$ to begin, but once initiated will propagate through the medium with
a speed comparable to speed of reconnection in the individual knots.  In a more
realistic situation,  the net effect will be a kind of  modified sandpile model
for magnetic reconnection in the solar corona and chromosphere.  As the density
of knots increases, and the energy available through magnetic reconnection
increases, the chance of a successfully propagating reconnection front will
increase.

\section{4. LV99 model and particle acceleration}

As any model of fast reconnection, LV99 model transfers only a small fraction of energy into heat via
Ohmic heating. Most of the energy ends up in the motion of magnetic fields and plasmas. Turbulent
outflow naturally suggests the possibility of the second order Fermi acceleration of energetic particles. 
However, this is not the only process of acceleration.

\begin{figure*}[!t]
\includegraphics[width=0.4\textwidth]{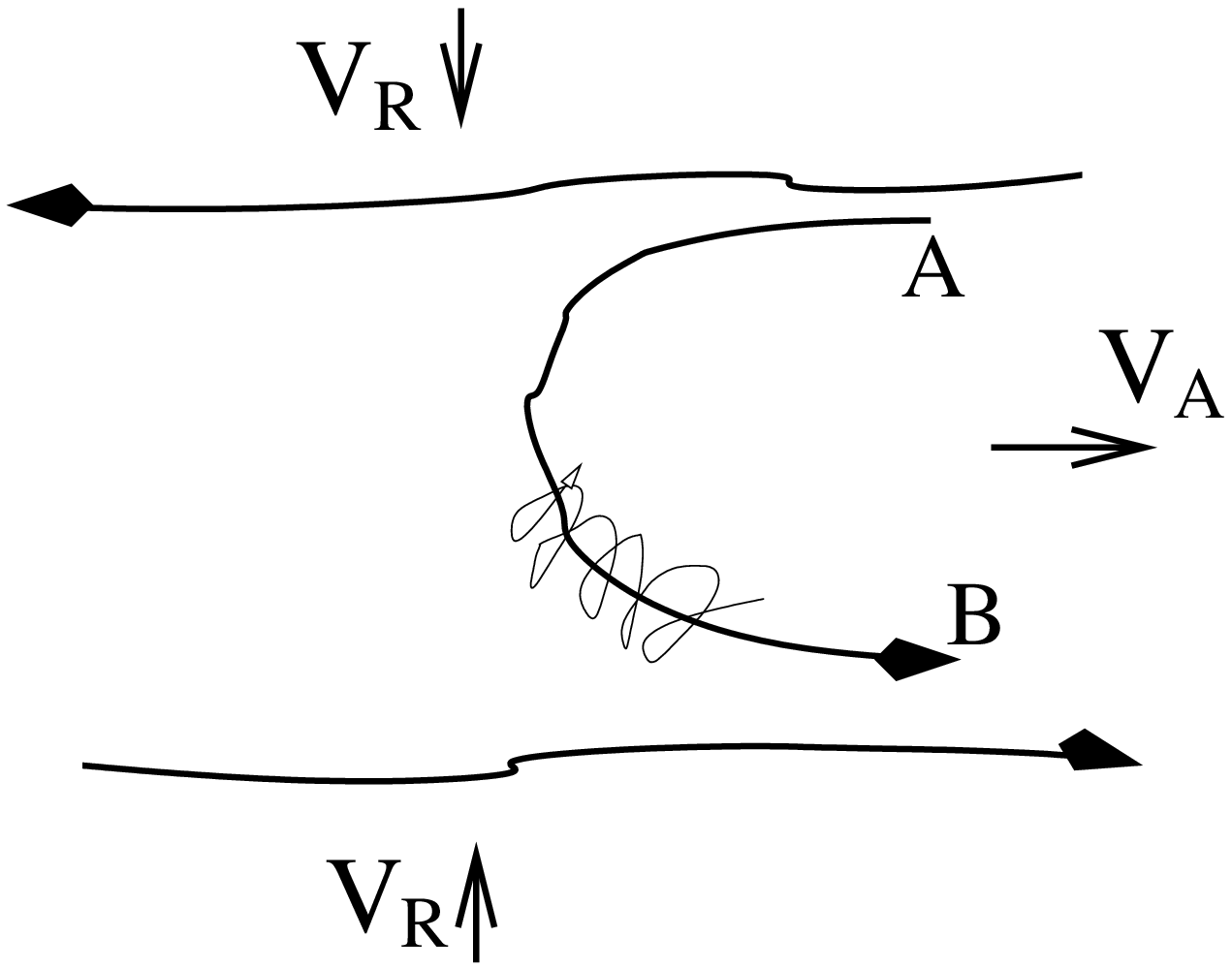}
\includegraphics[width=0.4\textwidth]{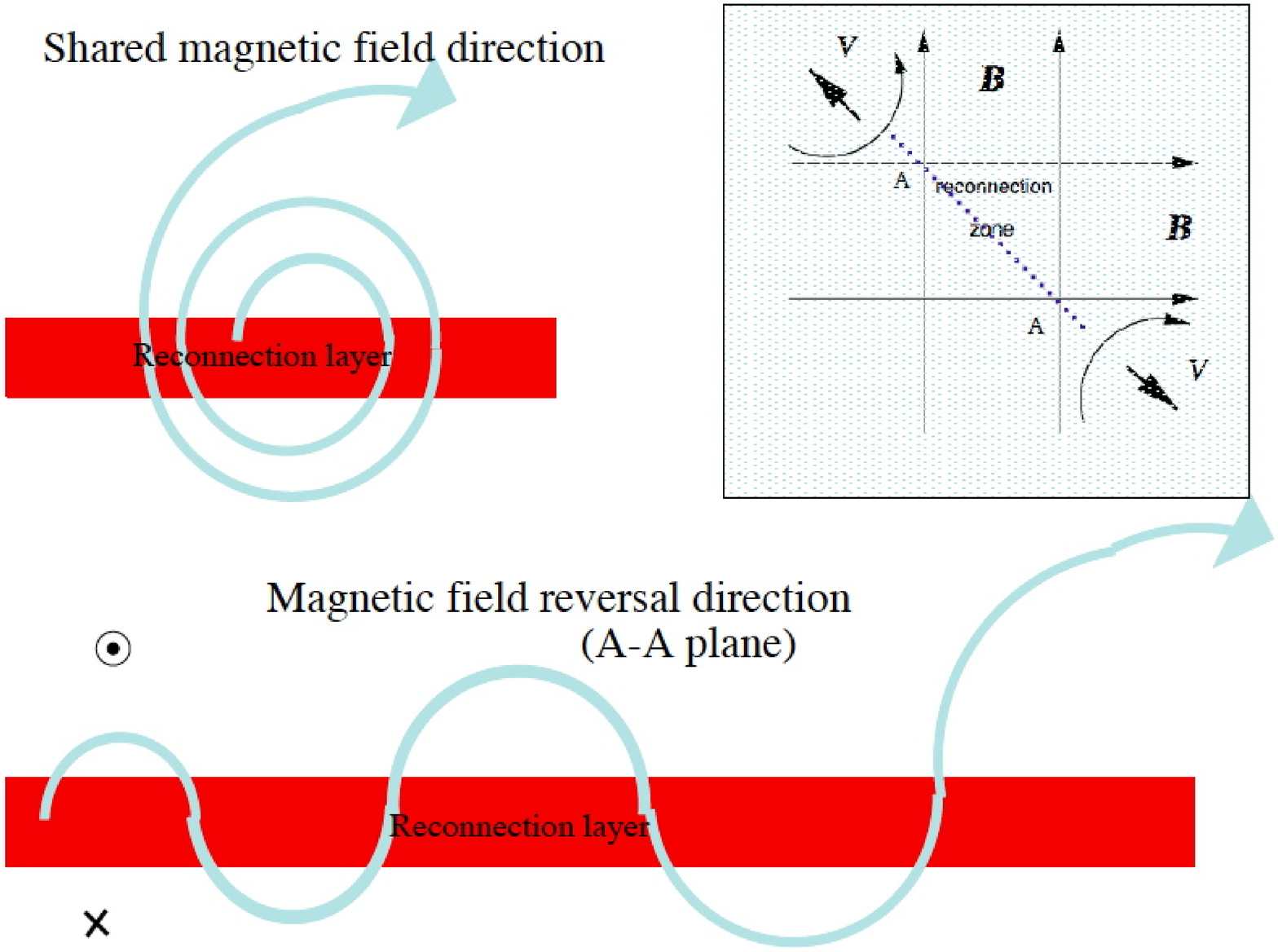}
\caption{{\it Left} Cosmic rays spiral about a reconnected magnetic field line and bounce
back at points A and B. The reconnected regions move towards each other with the
reconnection velocity $V_R$.   From Lazarian 2005. {\it Right}:  Particles with a large Larmor radius gyrate about the magnetic field
shared by two reconnecting fluxes (the latter is frequently referred to as
``guide field''. As the particle interacts with converging magnetized flow
corresponding to the reconnecting components of magnetic field, the particle
gets energy gain during every gyration.} 
\label{figure5}
\end{figure*}

Analyzing the dynamics of the volume reconnection that occurs within LV99 outflow zone, de Gouveia dal Pino
\& Lazarian (2005, henceforth GL05, see also Lazarian 2005) showed that the first order Fermi acceleration should take place
as the energetic particles bounce between approaching mirrors and follow the contracting loops of magnet
flux (see Figure \ref{figure5}). 

The acceleration of energetic particles in contracting loops shown in left plot of Figure \ref{figure5} is analogous to 
the acceleration proposed later in an influential paper by Drake et al. (2006). The difference is that the latter paper appealed to magnetic 
islands produced by collisionless reconnection. However, in any realistic 3D configuration of reconnecting magnetic
field, 2D islands are improbable and 3D magnetic loops should be considered instead. The difference between
the nature of LV99 and collisionless reconnection is not important as far as the reconnection process is concerned. 
Both processes increase the parallel component of the bouncing particle and difference between treatments in
GL05 and Drake et al. (2006) is that the latter paper took into account the possible effects of feedback via firehose
instability on the particle acceleration. 

\begin{figure*}
\includegraphics[width=0.8\textwidth]{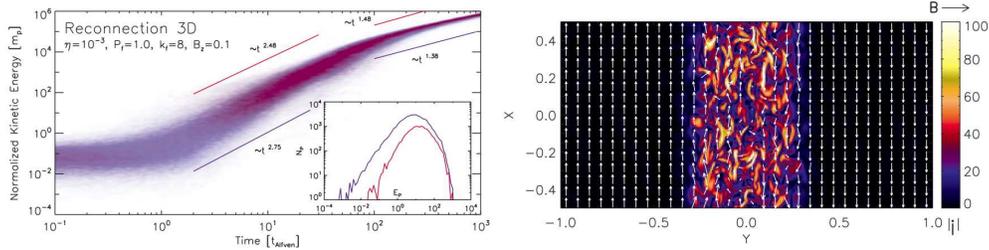}
\caption{ {\it Left} Acceleration of energetic particles in magnetic reconnection in the presence of turbulence. Red color corresponds to the
preferential acceleration in terms of the parallel momentum, blue corresponds to the perpendicular momentum increase. {\it Right}: Reconnection
layer corresponding to the process of the acceleration. From Kowal et al. 2012.
\label{figure6}}
\end{figure*}

In addition to the acceleration of terms of the parallel momentum, the first order Fermi acceleration in terms of perpendicular momentum
is also possible (see Figure \ref{figure5}, Right).  There the particle with a large Larmor radius is
bouncing back and forth between converging mirrors of reconnecting magnetic
field systematically getting an increase of the perpendicular component of its
momentum.  Both processes take place in reconnection layers. Both processes of the acceleration in terms of parallel and perpendicular momentum 
are observed in the numerical studies of magnetic reconnection in the presence of turbulence as it is shown in Kowal et al. (2012a). Figure \ref{figure6} illustrates the acceleration process that takes place in the turbulent reconnection layer. It is easy to see that both parallel and perpendicular momenta
increase in the process.

The first order acceleration arising from the reconnection of turbulent astrophysical magnetic fields was invoked to explain the origin of anomalous
cosmic rays in Lazarian \& Opher (2009) (see also Drake et al. 2010) and cosmic ray anisotropies in Lazarian \& Desiati (2010). A discussion of more
cases of the acceleration via reconnection is surely to follow. For instance, as we alluded earlier, the acceleration within the reconnection regions
that separate of spirals of pulsar winds is very promising. Reconnection is also invoked for the acceleration of cosmic rays in galaxy clusters 
(Brunetti \& Lazarian 2012). In some cases the acceleration is considered within loops of contracting magnetic field without direct reference to
turbulent reconnection (see Hoshino 2011). However, as turbulence is ubiquitous in astrophysical environments it
is difficult to avoid its crutial influence on reconnection (see Figure \ref{figure5}, right). Moreover, as we discussed in \S 3, in magnetically
dominated environments, i.e. in the environments where turbulent reconnection is important, the reconnection is bound to induce turbulence. 
 Below we discuss the energization of particles through reconnection that results in gamma ray bursts.

\section{5. Application to Gamma Ray Bursts}

\subsection{Problem of Gamma Ray Bursts}

New questions on Gamma Ray Bursts (GRBs) were proposed thanks to Swift and Fermi observations. In particular, recent Fermi observation of GRB 080916C indicates that the bright photosphere
emission associated with a putative fireball is missing, which is 
challenging to the traditional fireball internal shock (IS) model. A Poynting-flux-dominated outflow from the central engine is more preferable at least for this burst.

\begin{figure}
\centering
\includegraphics[width=0.48\columnwidth]{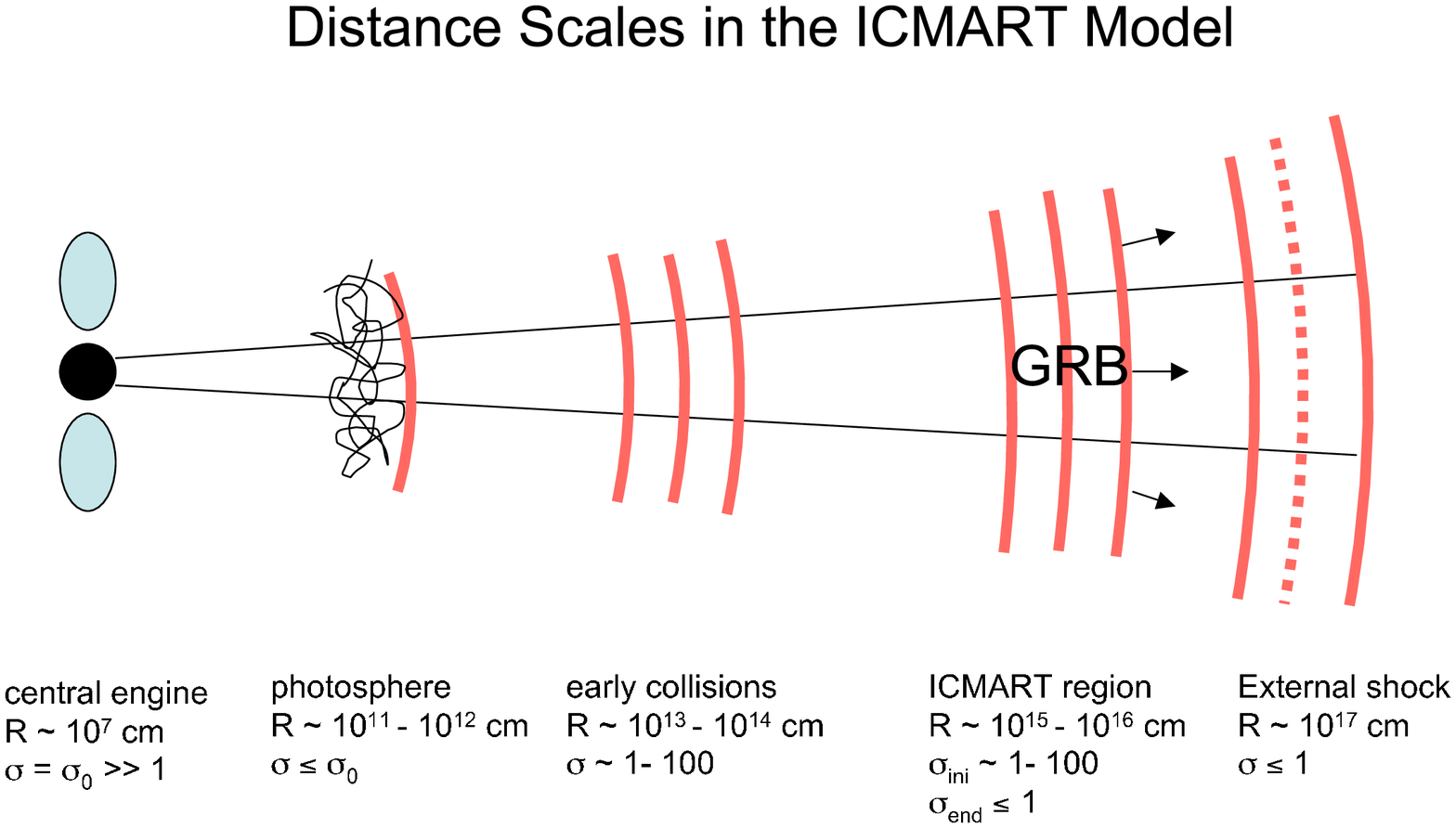}
\includegraphics[width=0.48\columnwidth]{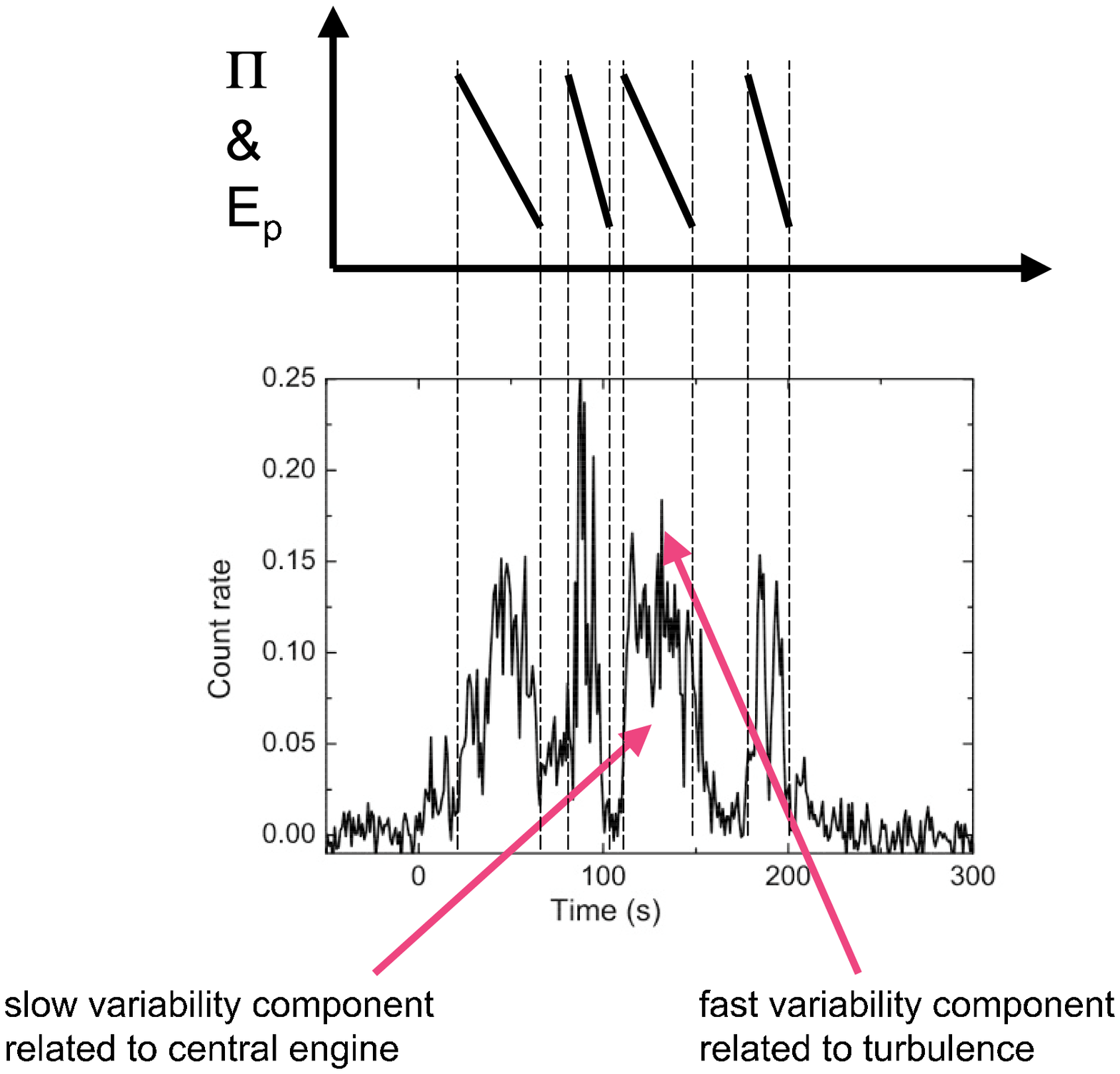}
\caption{\small {\it Left panel}: A cartoon picture of the ICMART model. The
typical distances and $\sigma$ values of various events are marked. {\it Right
panel}: An example of GRB light curve that shows two variability time scales.
The light curve of GRB 050607 is taken from the NASA Swift GRB archival data web
site http://swift.gsfc.nasa.gov/docs/swift/archive/grb\_table/
grb\_lookup.php?grb\_name=050607. The predictions of decreasing gamma-ray
polarization degree $\Pi$ and the spectral peak energy $E_p$ within individual
pulses are indicatively presented. Detailed decaying functions would be
different depending on the details of evolution of magnetic field configuration,
$\sigma$ value, as well as balance between heating and cooling of electrons. The
general decreasing trend is robust. (from Zhang \& Yan 2011)} \label{icmart}
\end{figure}

In the standard fireball IS scenario (Pac\'ynski et al. 1994; Shemi et al. 1990), magnetic fields are not important, i.e. $\sigma << 1$, where $\sigma$ is the ratio between the Poynting flux and the matter (baryonic $+$ leptonic) flux. An alternative picture is that the magnetic field is dynamically important
in GRB outflows, i.e.
$\sigma >> 1$. In this case GRB radiation would be powered by the magnetic field energy in the ejecta (see, e.g., Usov 1992;  Thompson 1994; Lazarian et al. 2003; Lyutikov \& Blandford 2003). 
The analogy with AGNs
jets also implies that magnetic dominated
jets as a viable option for the source of GRBs. It has become clear that the inner engine cannot accelerate
baryonic dominated jets for AGNs, suggesting that similarly GRBs are Poynting flux dominated at least in its inner
region the jet.

\subsection{LV99 and GRBs}

Magnetic reconnection has been suggested as the engine of GRBs. The problem lay, however, in the intrinsic difficulty of
reconnection since it is a very slow process in ordered fields. As with the case
for solar flares, both a slow phase and a fast bursty phase are required for reconnection. Essential progress was made by Lazarian et al. (2003), where a new GRB model of self-adjusted
reconnection was proposed based on the LV99 reconnection\footnote{Indeed LV99 was proposed as a model of reconnection in non-relativistic case. However, the scaling and anisotropies
of magnetized turbulence were found to be the same for both relativistic
and non-relativistic cases (Cho \& Lazarian 2012). Since turbulence properties
are the most important ingredient that determines the reconnection rate in LV99
model, it is natural to assume that the reconnection is similar in relativistic and
non-relativistic cases. In the former case, we expect the reconnection velocity 
to approach the velocity of light, as $V_A\rightarrow c$.}. As the turbulence builds up the
reconnection rate increases, which induces a positive feedback, resulting in the
explosive reconnection. This can be an alternative scenario to
gamma ray bursts (Lazarian et al. 2003). More recently the model was elaborated
and connected with observational data in (Zhang \& Yan 2011).

As in the IS model, the mini-shells interact internally at the
radius $R_{\rm IS}\sim \Gamma^2 c \Delta t$.  Most of these early collisions,
however, do not trigger any magnetic reconnection, instead only lead to perturbations of the magnetic field in the ejecta. As in other astrophysical environment,
the system is prone to turbulence because of high Reynolds and magnetic Reynolds
numbers\footnote{This is a nontrivial statement since the system
is highly collisionless. Nonetheless, due to the high
magnetization, most kinetic motions are perpendicular to
magnetic field so that it is the much suppressed perpendicular viscosity and
resistivity that should be adopted.} Since GRBs outflow is highly
relativistic, the magnetic field is unlikely to be totally distorted. In any
case, fast reconnection can be triggered in weakly stochastic magnetic field
(LV99). The details require further studies especially in the regime of
relativistic turbulence.

The perturbations accumulate as the mini-shells propagate outward (see
Fig.\ref{icmart} {\em right}) till a point where the turbulence reaches the threshold to trigger a fast reconnection. Fast reconnection events rapidly eject outflows, which makes the system even more  turbulence. This leads to a run-away release of the magnetic field energy in a reconnection/turbulence avalanche. This is one ICMART event, corresponding to one GRB pulse. The $\sigma$ value decreases to order of unity from the original value during this process.

A GRB consists of several ICMART events (i.e. broad pulses). The peak energy
$E_p$ is expected to drop across each pulse. The $\gamma$-ray
polarization degree is also expected to fall from $\sim 50-60\%$ to $\sim$ a few
$\%$ during each ICMART event. There should be two variability components in the GRB
light curves , one slow component
associated with the central engine activity, and a fast component related
to the relativistic magnetic turbulence  (see Fig.\ref{icmart}). More details can be found in Zhang \& Yan (2011).

\section{6. Discussion and Summary}

Turbulence is a natural state of astrophysical fluids that are characterized by very high Reynolds numbers.
Therefore it should be taken into account for various astrophysical processes and magnetic reconnection
is not an exception. In fact, in particular cases when the level of turbulence in the astrophysical systems is low,
{\bf which} can be the case of low $\beta$ plasmas, the reconnection itself induces turbulence resulting in 
reconnection instability and bursts of reconnection as we discussed in \S 3.

LV99 model provides analytical predictions for the rate of reconnection in the presence of magnetic field and those
predictions have been successfully tested numerically in Kowal et al. (2009, 2012a). In addition, the necessity of the
 LV99 reconnection for explaining the properties of MHD turbulence and making MHD turbulence self-consistent
 was established both in LV99 and ELV11. The model can also account for observed properties of solar flares, e.g.
 bursts of reconnection, triggering of reconnection, fast reconnection in collisional media etc. This makes it 
 advantageous to apply the model to describe reconnection in various astrophysical contexts. 
 
 The energy released within LV99 reconnection can efficiently accelerate particles via both the first order Fermi acceleration
 and the second order Fermi acceleration. This can power gamma ray busts. The observations are consistent with the 
 theoretical predictions and this calls for more elaborate modeling of reconnection associated with the gamma ray bursts.

\begin{theacknowledgments}
 A.L. research is supported by the NASA Grant NNX09AH78G, as well as the support
of the NSF Center for Magnetic Self-Organization.  The Humboldt Award at the
Universities of Cologne and Bochum, as well as Vilas Associate Award are acknowledged. H. Y. acknowledges the support from NSFC Grant AST -11073004 and the visiting professorship from IIP, Natal, Brazil.
\end{theacknowledgments}


\begin{thebibliography}{9}

\bibitem[Bhattacharjee et al.(2009)]{2009PhPl...16k2102B} Bhattacharjee, 
A., Huang, Y.-M., Yang, H., 
\& Rogers, B.\ 2009, Physics of Plasmas, 16, 112102

\bibitem{bisc} Biskamp, D. 2003, Magnetohydrodynamic Turbulence, CUP

\bibitem{Brunetti} Brunetti, G. \& Lazarian, A. 2012, MNRAS, submitted

\bibitem{cho} Cho, J. \& Lazarian, A. 2012, ApJ, in preparation

\bibitem[Ciaravella \& Raymond(2008)]{CiaravellaRaymond08} Ciaravella, A., \& Raymond, J.~C.\ 2008, ApJ, 686, 1372

\bibitem[de Gouveia dal Pino 
\& Lazarian(2005)]{2005A&A...441..845D} de Gouveia dal Pino, E.~M., \& Lazarian, A.\ 2005, \aap, 441, 845

\bibitem[Drake et al.(2006)]{Drakeetal06} Drake, J.~F., Swisdak, 
M., Che, H., \& Shay, M.~A.\ 2006, Nature, 443, 553

\bibitem[Drake et al.(2010)]{2010ApJ...709..963D} Drake, J.~F., Opher, M., 
Swisdak, M., \& Chamoun, J.~N.\ 2010, ApJ, 709, 963

\bibitem{Eyink11} Eyink, G. L.\ 2011,Phys. Rev. E 83, 056405

\bibitem[Eyink et al.(2011)]{2011ApJ...743...51E} Eyink, G.~L., Lazarian, 
A., \& Vishniac, E.~T.\ 2011, ApJ, 743, 51 (ELV11)

\bibitem[Jacobson 
\& Moses(1984)]{Jacobson84} Jacobson, A.~R., \& Moses, R.~W.\ 1984, \pra, 29, 3335

\bibitem[Jokipii (1973)]{Jokipii73}
Jokipii, J.~R.\ 1973, 
ApJ, 183, 1029. 

 \bibitem{go} Goldreich, P. \& Sridhar, S. 1995, ApJ, 438, 763
 
 \bibitem{} Kowal, G., de Gouveia Dal Pino, E., \& Lazarian, A. 2012, Phys. Rev. Lett. 
 
\bibitem[Kowal et al.(2012)]{2012NPGeo..19..297K} Kowal, G., Lazarian, A., 
Vishniac, E.~T., 
\& Otmianowska-Mazur, K.\ 2012, Nonlinear Processes in Geophysics, 19, 297 
 
 \bibitem{Kowaletal09} Kowal, G., Lazarian, A., Vishniac, E.~T., \& Otmianowska-Mazur, K.\ 2009, ApJ, 700, 63
 
 \bibitem[Lapenta (2008)]{Lapenta08}
Lapenta, G.\ 2008, 
Phys. Rev. Lett., 100, 235001

\bibitem[Lazarian(2005)]{2005AIPC..784...42L} Lazarian, A.\ 2005, Magnetic 
Fields in the Universe: From Laboratory and Stars to Primordial 
Structures., 784, 42 

\bibitem[Lazarian(2006)]{lazarian2006} Lazarian, A.\ 2006, ApJL,
645, L25

\bibitem[Lazarian et
al.(2011)]{2011P&SS...59..537L} Lazarian, A., Kowal, G., Vishniac, E., \& de Gouveia Dal Pino, E.\ 2011, Planetary and Space Science, 59, 537


\bibitem[Lazarian 
\& Desiati(2010)]{2010ApJ...722..188L} Lazarian, A., \& Desiati, P.\ 2010, ApJ, 722, 188

\bibitem[Lazarian 
\& Opher(2009)]{2009ApJ...703....8L} Lazarian, A., \& Opher, M.\ 2009, ApJ, 703, 8

\bibitem[Lazarian 
\& Pogosyan(2004)]{2004ApJ...616..943L} Lazarian, A., \& Pogosyan, D.\ 2004, ApJ, 616, 943

\bibitem[Lazarian 
\& Vishniac(2009)]{2009RMxAC..36...81L} Lazarian, A., \& Vishniac, E.~T.\ 2009, Revista Mexicana de Astronomia y Astrofisica Conference Series, 36, 81

 \bibitem{la5} Lazarian, A. \& Vishniac, E. 1999, ApJ, 517, 700 (LV99)

\bibitem{la4} Lazarian, A., Vishniac, E. \& Cho, J. 2004, \apj, 603, 180

\bibitem{l12} Lazarian, A., Vlahos, L., Kowal, G., Yan, H., Beresnyak, A., de Gouveia Dal Pino E. 2012, Space Science Review, DOI 10.1007/S 11214-9934-7

 
 \bibitem[Loureiro et al.(2007)]{2007PhPl...14j0703L} Loureiro, N.~F., 
Schekochihin, A.~A., \& Cowley, S.~C.\ 2007, Physics of Plasmas, 14, 100703

\bibitem[Matthaeus 
\& Lamkin(1985)]{1985PhFl...28..303M} Matthaeus, W.~H., \& Lamkin, S.~L.\ 1985, Physics of Fluids, 28, 303 

\bibitem[Matthaeus 
\& Lamkin(1986)]{1986PhFl...29.2513M} Matthaeus, W.~H., \& Lamkin, S.~L.\ 1986, Physics of Fluids, 29, 2513

 
\bibitem[\protect\citeauthoryear{{Pacz\'ynski} and {Xu}}{{Pacz\'ynski} and
  {Xu}}{1994}]{paczynski94}
{Pacz\'ynski}, B., \& {Xu}, G. 1994, \apj,~427, 708.

\bibitem[Petschek(1964)]{25} Petschek, H.E. Magnetic field annihilation. The Physics of Solar Flares, AAS-NASA Symposium (NASA SP-50), ed. WH. Hess (Greenbelt, MD: NASA) 425

\bibitem[Shay \& Drake(1998)]{sha98} Shay, M.~A. \& Drake, J.~F., 1998, Geophys. Res. Letters Geophysical Research Letters, 25, 3759-3762

\bibitem[\protect\citeauthoryear{{Shemi} and {Piran}}{{Shemi} and
  {Piran}}{1990}]{shemi90}
{Shemi}, A., \& {Piran}, T. 1990, \apjl,~365, L55.

\bibitem[Speiser(1970)]{Speiser70} Speiser, T.~W.\ 1970, \planss, 18, 613

\bibitem[Sych et 
al.(2009)]{Sychetal09} Sych, R., Nakariakov, V.~M., Karlicky, M., 
\& Anfinogentov, S.\ 2009, \aap, 505, 791

\bibitem[\protect\citeauthoryear{{Thompson}}{{Thompson}}{1994}]{thompson94}
{Thompson}, C. 1994, \mnras,~270, 480.

\bibitem[\protect\citeauthoryear{{Usov}}{{Usov}}{1992}]{usov92}
{Usov}, V.~V. 1992, Nature,~357, 472.

\bibitem[Uzdensky 
\& Kulsrud(2006)]{UzdenskyKuslrud06} Uzdensky, D.~A. 
\& Kulsrud, R.~M.\ 2006,  Physics of Plasmas, 13, 062305

\bibitem[Uzdensky et al. (2010)]{Uzdenskyetal10}
Uzdensky, D.~A., Loureiro, N.~F. and Schekochihin, A.~A.\ 2010. 
\prl 105, 235002.


\bibitem[]{ZhangYan}Zhang, B. \& Yan, H. 2011, ApJ, 726, 90
\end{thebibliography}
\end{document}